\documentclass{article}
\usepackage{amsmath,graphicx,mlspconf}

\usepackage{url}
\usepackage{comment}
\usepackage{booktabs}
\usepackage{xcolor}
\usepackage{colortbl}
\usepackage{subcaption}
\usepackage{graphicx}
\usepackage{arydshln}
\usepackage{multirow}
\usepackage{amsmath}
\usepackage{pifont}

\usepackage{amssymb}
\usepackage{pifont}

\definecolor{almondlow}{RGB}{252,239,219} 
\definecolor{almondmiddle}{RGB}{237,225,206} 
\definecolor{almondhigh}{RGB}{224,213,194} 
\definecolor{almondultra}{RGB}{214,204,186} 
\definecolor{melon}{RGB}{232,156,124} 
\definecolor{champagne}{rgb}{0.97, 0.91, 0.81}
\definecolor{baselinecolor}{rgb}{1,1,1}
\definecolor{corn}{rgb}{0.98, 0.93, 0.36}
\definecolor{teagreen}{rgb}{0.82, 0.94, 0.75}
\definecolor{predcolor}{rgb}{0.74, 0.83, 0.9}
\definecolor{decorrcolor}{rgb}{0.98, 0.85, 0.87} 
\definecolor{camouflagegreen}{rgb}{0.47, 0.53, 0.42}
\definecolor{brightube}{rgb}{0.82, 0.62, 0.91} 
\definecolor{paletaupe}{rgb}{0.74, 0.6, 0.49}
\definecolor{pastelviolet}{rgb}{0.8, 0.7, 0.79}
\definecolor{knowcolor}{rgb}{0.80, 0.94, 0.75}
\definecolor{offlinecolor}{rgb}{1,1,1}
\definecolor{firsttaskcolor}{rgb}{0.97, 0.97, 0.97}
\definecolor{azure(colorwheel)}{rgb}{0.0, 0.5, 1.0}
\definecolor{gray(x11gray)}{rgb}{0.75, 0.75, 0.75}
\definecolor{lightgray}{rgb}{0.90, 0.90, 0.90}
\definecolor{darkgray}{rgb}{0.66, 0.66, 0.66}
\definecolor{pastelorange}{rgb}{1.0, 0.7, 0.28}
\definecolor{pastelyellow}{rgb}{0.99, 0.99, 0.59}
\definecolor{pistachio}{rgb}{0.58, 0.77, 0.45}
\definecolor{shamrockgreen}{rgb}{0.0, 0.62, 0.38}
\definecolor{pastelred}{rgb}{1.0, 0.41, 0.38}
\definecolor{slateblue}{rgb}{0.42, 0.35, 0.8}
\definecolor{pinksecondbest}{RGB}{252, 241, 241}
\definecolor{aqua}{rgb}{0.0, 1.0, 1.0}
\definecolor{aquamarine}{rgb}{0.5, 1.0, 0.83}

\usepackage{graphicx}
\usepackage{soul}

\usepackage{tcolorbox}

\toappear{2024 IEEE International Workshop on Machine Learning for Signal Processing, Sept.\ 22--25, 2024, London, UK}

\title{Parameter-Efficient Transfer Learning of Audio
Spectrogram Transformers}
\name{
    Umberto Cappellazzo$^{\heartsuit}$
    \qquad Daniele Falavigna$^{\clubsuit}$
    \qquad Alessio Brutti$^{\clubsuit}$
    \qquad Mirco Ravanelli$^{\spadesuit}$
}
\address{
    $^{\heartsuit}$ University of Trento, Italy
    $^{\clubsuit}$ Fondazione Bruno Kessler, Trento 
    $^{\spadesuit}$ Concordia University, Canada  \\
}

\begin{document}

\maketitle

\begin{abstract}
Parameter-efficient transfer learning (PETL) methods have emerged as a solid alternative to the standard full fine-tuning approach. They only train a few extra parameters for each downstream task, without sacrificing performance and dispensing with the issue of storing a copy of the pre-trained model for each task. For audio classification tasks, the Audio Spectrogram Transformer (AST) model shows impressive results. However, surprisingly, how to efficiently adapt it to several downstream tasks has not been tackled before. In this paper, we bridge this gap and present a detailed investigation of common PETL methods for the adaptation of the AST model to audio/speech tasks. Furthermore, we propose a new adapter design that exploits the convolution module of the Conformer model, leading to superior performance over the standard PETL approaches and surpassing or achieving performance parity with full fine-tuning by updating only 0.29\% of the parameters. Finally, we provide ablation studies revealing that our proposed adapter: 1) proves to be effective in few-shot efficient transfer learning, 2) attains optimal results regardless of the amount of the allocated parameters, and 3) can be applied to other pre-trained models. Our code is available at \url{https://github.com/umbertocappellazzo/PETL_AST}.

\end{abstract}

\section{Introduction}
Leveraging large pre-trained models for downstream tasks has become a cornerstone of several machine learning domains like natural language processing (NLP) and audio/speech processing. The typical paradigm involves adapting the whole model to each downstream task \cite{lv2023full, wang2021fine} (i.e., full fine-tuning). Despite achieving remarkable results, this approach leads to a specialized model for each task, which is unfeasible when fine-tuning a model on numerous downstream tasks. 

To alleviate this issue, the research community is increasingly focusing on parameter-efficient transfer-learning (PETL) methods, whereby only a small amount of extra parameters is learned for each task while keeping the pre-trained model frozen \cite{xu2023parameter, he2021towards, lialin2023scaling}. In doing so, the risk of catastrophic forgetting the pre-trained model's knowledge is also highly reduced, a common problem in continual learning scenarios \cite{cappellazzo1, cappellazzo2}. 
For example, \textbf{prompt-tuning} methods insert trainable continuous vectors in the input or hidden state of the model, known as prompts \cite{lester2021, jia2022visual}. Alternatively, low-rank modules called \textbf{adapters}, which follow a bottleneck architecture with a very small intermediate dimension, are introduced into each layer. 
Another popular method, \textbf{LoRA} (Low-Rank Adaptation), leverages low-rank matrix decomposition of pre-trained weight matrices \cite{hu2021lora}. Several variants of LoRA have been recently proposed to enhance the original implementation leading to better performance and stability \cite{zhang2023adaptive, liu2024dora}.

Recently, PETL methods have garnered much attention also in the audio and speech fields. For example, \cite{lin2024peft, chen2023exploring, li2023evaluating} provide extensive experiments on the use of PETL approaches and their combination for self-supervised learning speech models. Also for automatic speech recognition adapters have proven to be an effective solution \cite{kessler2022adapter, tomanek-etal-2021-residual}. 
For audio classification, the Audio Spectrogram Transformer (AST) \cite{gong2021ast} obtains superb results, standing out as the state-of-the-art model for several downstream tasks. The problem of how to efficiently transfer the knowledge of the AST is of crucial importance, especially given the typical computational and storage constraints of audio devices. \textit{Surprisingly, this topic has received minimal attention} \cite{adapter_cl}. Therefore, driven by 1) the absence of previous works, 2) the excellent results obtained by PETL methods in different domains for transformer models, and 3) the need to efficiently adapt the AST model to several downstream tasks, we ask the following question: 
\begin{tcolorbox}
\textbf{(}$\mathbf{Q}$\textbf{)} \textit{Can we exploit state-of-the-art PETL methods for the efficient fine-tuning of AST to audio/speech downstream tasks?}
\end{tcolorbox}

We methodically investigate this research question \textbf{(}$\mathbf{Q}$\textbf{)}, and to do so we provide a framework whereby we can study the performance attained by several PETL methods on five audio/speech benchmarks. Furthermore, from our experiments, we notice that the bottleneck adapter struggles to achieve on-par performance with respect to the full fine-tuning approach for speech tasks. We conjecture that this is attributable to the overly simplistic design of the bottleneck adapter, where only linear layers are adopted, which hinders a complete learning of the task at hand. As a consequence, we propose a new adapter design that hinges upon the convolution module of the Conformer model. Our proposed \textit{Conformer adapter} highly benefits from the introduction of the depthwise convolution layer, which allows not only to capture local spatial correlations but also trim down the number of parameters, thus bridging the gap with the full fine-tuning method. 

We carry out extensive experiments leading to multiple findings: \ding{182} among the standard PETL methods, \textit{LoRA} and \textit{Houlsby bottleneck adapter} achieve the best performance overall, with \textit{LoRA} using fewer parameters; \ding{183} our proposed \textbf{conformer adapter} provides considerable improvements over the bottleneck adapter, surpassing or attaining performance parity with respect to the full fine-tuning approach while using only $0.29/0.59\%$  parameters compared to it for the Pfeiffer/Houlsby configuration; \ding{184} we study the PETL methods under few-shot settings and their scalability with respect to the number of trainable parameters, validating the efficacy of our proposed adapter; \ding{185} we show empirically that the kernel size of the depthwise convolution is a key parameter to attain the best performance; \ding{186} we finally show that the conformer adapter can be also harnessed for the efficient fine-tuning of another pre-trained model like Wav2Vec 2.0. 

\begin{figure}[t]
    \centering
    \includegraphics[width=8cm]{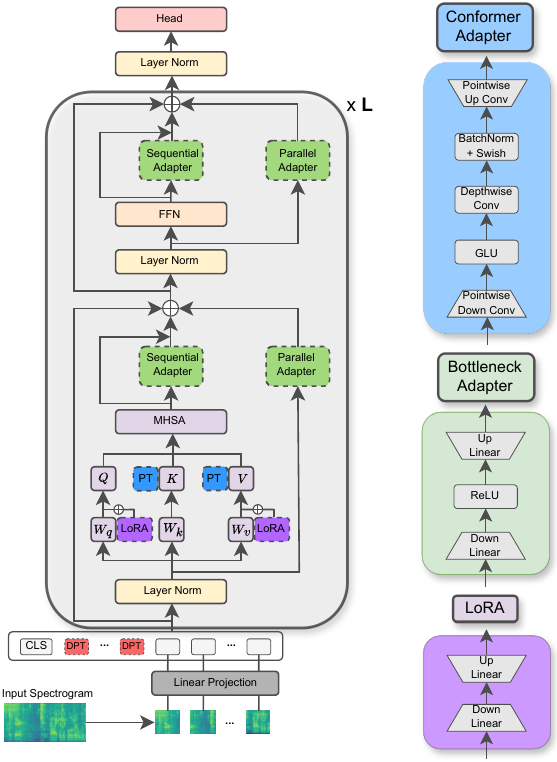}
    \caption{\textbf{Left}: illustration of the AST model and the integration of PETL methods into it. We use blocks with dashed outlines to characterize the added modules by those methods. \textbf{Right}: the inner structure of LoRA, Bottleneck adapter and our proposed Conformer adapter.}
    \label{fig:main_diagram}
\end{figure}

\section{Methodology}
\subsection{AST Model Recap}
The Audio Spectrogram Transformer (AST) is an attention-based model that achieves state-of-the-art results on various audio and speech tasks \cite{gong2021ast, ssast}. The AST model receives as input audio spectrograms that are patchified and then a linear projection is applied to each patch. This results in a sequence of $N$ tokens of size $d = 768$, which we refer to as $\textbf{X}_{in} \in \mathbb{R}^{N \times d}$. AST comprises $12$ attention layers, each of which is composed of two sub-layers: a \textit{multi-head self-attention} (MHSA) sub-layer and a fully-connected feed-forward (FF) sub-layer. The output of each transformer layer, $\textbf{X}_{out} \in \mathbb{R}^{N \times d}$ (we omit for simplicity the index of the layer), is computed as follows: 
\begin{equation}
    \mathbf{X}_{out}= \hat{\mathbf{X}} + \text{FF}(\text{LN}(\hat{\mathbf{X}})), \hat{\mathbf{X}} = \mathbf{X}_{in} + \text{MHSA}(\text{LN}(\mathbf{X}_{in})).
    \label{eq:1}
\end{equation}
Both blocks, MHSA and FFN, include residual connections and layer normalizations (LN) \cite{LN}, with the LN applied within the residual branch (i.e., Pre-LN). 

\subsection{Overview of Parameter-efficient Transfer Learning Methods}
We now introduce the PETL techniques we used in our experiments: LoRA, prompt/prefix-tuning, and adapter-tuning.

\textbf{LoRA \cite{hu2021lora}}. LoRA introduces trainable low-rank matrices into transformer layers to approximate the weight updates. For a pre-trained weight matrix $\mathbf{W} \in \mathbb{R}^{d \times d_k}$, LoRA represents its update with a low-rank decomposition $\mathbf{W}+ \Delta W =\mathbf{W} + \mathbf{A}\mathbf{B}$, where $\mathbf{A} \in \mathbb{R}^{d \times r}$, $\mathbf{B} \in \mathbb{R}^{r \times d}$ are learnable and $r << d$. LoRA typically applies this update to the query and value projection matrices, $\mathbf{W}_q$ and $\mathbf{W}_v$, in the MHSA sub-layer. LoRA computes the query and value matrices like this:
\begin{equation}
    \mathbf{Q} / \mathbf{V} =  \mathbf{X}_{in}\mathbf{W}_{q/v} + s \cdot \mathbf{X}_{in} \mathbf{A}_{q/v} \mathbf{B}_{q/v},
\end{equation}
where $s$ is a tunable scalar hyperparameter.

\textbf{Prefix-tuning/Prompt-tuning \cite{li2021prefix, lester2021}}. Prefix-tuning \cite{li2021prefix} inserts $p$ learnable continuous embeddings of dimension $d$ (i.e., \textit{prompts}) to the keys and values of the MHSA block at every layer. Prompt-tuning \cite{lester2021, jia2022visual}, instead, prepends the prompts in the input space after the projection layer. Following \cite{jia2022visual}, we consider the ``\textit{shallow}'' prompt-tuning version (SPT) where all the prompts are prepended to the first transformer layer, and the ``\textit{deep}'' version (DPT) by prepending the prompts uniformly to each transformer layer.

\textbf{Bottleneck Adapter \cite{houlsby2019parameter, pfeiffer2020adapterfusion}}. Adapters are light subnetworks that are inserted into every transformer layer. To keep the number of parameters limited, adapters exploit a \textit{bottleneck} architecture. The input sequence of hidden dimension $d$ is first down-projected (parametrized by $\mathbf{W}_{down}) \in \mathbb{R}^{d \times r}$ into a low-dimensional space with size $r$ (the bottleneck dimension), followed by a non-linear activation function $f(\cdot)$ (e.g., ReLU), and then up-projected back to the original dimension d ($\mathbf{W}_{up} \in \mathbb{R}^{r \times d}$). We refer to this design as \textbf{bottleneck} adapter and it is the established choice in the NLP domain \cite{he2021towards, houlsby2019parameter}. Adapter-tuning is a flexible approach in that we can identify multiple ways in which an adapter can be included in a transformer layer, resulting in different configurations. For example, the adapter can be inserted only after the FF block, (\textit{Pfeiffer} \cite{pfeiffer2020adapterfusion}), or after both the MHSA and FF blocks (\textit{Houlsby} \cite{houlsby2019parameter}). Furthermore, the adapter can be included \textit{sequentially}, either after the FF block \cite{houlsby2019parameter} (sequential Pfeiffer) or after both FF and MHSA blocks \cite{mahabadi2021parameter} (sequential Houlsby), or \textit{parallel} to only the FFN block \cite{he2021towards, chen2022adaptformer}, or parallel to both FFN and MHSA blocks \cite{convpass}. Mathematically, if we consider, as an example, the \textit{Pfeiffer} configuration in which the Bottleneck adapter is placed \textit{sequentially} after the FF block and we let $\mathbf{X}_{\text{FF}} = \text{FF}(\text{LN}(\hat{\textbf{X}}))$, following the notation in Eq. \ref{eq:1}, the output is: 
\begin{equation}
\mathbf{X}_{out} = \hat{\mathbf{X}} + \mathbf{X}_{\text{FF}} + f(\hat{\mathbf{X}}\mathbf{W}_{down})\mathbf{W}_{up}. 
\end{equation}


\textbf{Conformer Adapter (Ours)}. As we will show in Section \ref{sec:main_results}, the bottleneck adapter attains competitive results for audio classification tasks, whereas for speech tasks the gap with the full fine-tuning approach is sizeable. We speculate that this happens because the linear design of the bottleneck adapter is not sufficient to disentangle the task at hand. For this reason, we propose to leverage the key block of the Conformer \cite{conformer}, a bleeding-edge model for several speech processing tasks: the \textit{convolution module}. This module highly relies on the depthwise convolution, which is appealing for our PETL setting for two main reasons: 1) it is used for capturing spatial correlations, a crucial aspect for speech downstream tasks, which the bottleneck adapter fails to accomplish, and 2) compared to a standard convolution, it requires fewer parameters, thus making it suitable for parameter-efficient adapters. Therefore, we propose a new adapter that uses the convolution module as the building block and we call it \textit{conformer adapter} (see Fig.~\ref{fig:main_diagram}, right). Specifically, the first pointwise convolution down-projects the input sequence to a dimension of $2r$. Then, the Gated Linear Unit (GLU) halves the hidden dimension to the bottleneck one, $r$. At this point, the intermediate sequence undergoes the depthwise convolution layer with kernel size equal to $k$ (refer to Section \ref{sec:ablations} for the analysis on this hyper-parameter), as well as the Batch Normalization and Swish activation. Finally, the dimension of the sequence is up-projected to the original $d$ through a pointwise convolution. We show the effectiveness of our conformer adapter in Section~\ref{sec:experiments}.

\section{Experiments}
\label{sec:experiments}

\subsection{Implementation Details}

\textbf{Datasets}. We evaluate the PETL methods on four audio/speech downstream classification tasks. (1) \textbf{Audio classification}: we use the ESC-50 and UrbanSound8K (US8K) datasets. ESC-50 (ESC) \cite{piczak2015esc} consists of $2,000$ 5-second-long environmental audio recordings of $50$ classes. US8K \cite{salamon2014dataset} includes $8,732$ labeled sound excerpts of urban sounds from $10$ classes. (2) \textbf{Keyword spotting}: Speech Commands V2 (GSC) \cite{warden2018speech} has $105,829$ 1-sec recordings of $35$ speech commands. (3) \textbf{Intent classification}: Fluent Speech Commands (FSC) \cite{lugosch2019speech} includes $30,043$ English utterances spanning $31$ classes. (4) \textbf{Emotion Recognition}: IEMOCAP (IEM) \cite{iemocap} comprises $10,039$ utterances
from 10 distinct speakers with $4$ emotional classes: \textit{neutral}, \textit{happy}, \textit{sad}, \textit{angry}.

\textbf{Baselines}. We include the \textit{full fine-tuning} method (FFT), which finetunes the full pre-trained AST model; and \textit{linear probing}, which only fine-tunes the classification head. We then study various PETL methods: shallow prompt-tuning (SPT), deep prompt-tuning (DPT), prefix-tuning (Pref-T), and BitFit \cite{zaken2021bitfit}, which is a common baseline that fine-tunes only the bias terms of the pre-trained
backbone. SPT adds all the $300$ prompts to the input of the first transformer layer, whereas DPT adds $25$ prompts to each transformer layer. We then include LoRA and bottleneck and conformer (ours) adapters. The dimension of the intermediate space for adapters and LoRA is $r = d/\text{RR}$, where $d = 768$ is the hidden dimension of the AST model and RR is the reduction rate. Unless otherwise stated, $r$ is set to $12$, $8$, and $6$ for bottleneck adapter, conformer, and LoRA, respectively. In this way, the resulting number of parameters is roughly the same. For LoRA, following \cite{hu2021lora}, the scaling factor is set to $s = \alpha/\text{RR}$, where $\alpha$ = 16 leads to the best results (i.e., $s = 8$). We also note that each adapter module is added in parallel to only the MHSA layer (Pfeiffer) or both the MHSA and FF layers (Houslby). For this reason, Houlsby adapters require twice as many parameters as Pfeiffer. Inserting the adapters sequentially leads to slightly worse results, yet we do not include these results for lack of space. Finally, for the speech tasks we set the kernel size of the depthwise convolution layer to 31, which is the original value proposed in \cite{conformer}, while for audio tasks we found that $k = 8$ gives the best results (we refer the reader to Section \ref{sec:ablations} for a detailed analysis).

\begingroup
\setlength{\tabcolsep}{3.3pt}
\newcolumntype{K}{!{\color{white}\ }c}

\begin{table}[t]
\centering
\caption{Performance evaluations of the PETL methods on $4$ datasets for AST. Best and second-best 
performances for each dataset are coloured in \sethlcolor{teagreen} \hl{\textbf{Green}} and \sethlcolor{pinksecondbest} \hl{Red}, respectively.}
\label{tab:main}
\begin{tabular}{lKKKKKK}
\toprule
\textbf{Method} & \textbf{Par} &\cellcolor{pastelyellow} \textbf{ESC} &\cellcolor{lightgray} \textbf{US8K} &\cellcolor{aquamarine} \textbf{GSC} &\cellcolor{melon} \textbf{FSC} & \cellcolor{champagne}\textbf{Avg}\\
\midrule
\textcolor{darkgray}{FFT} & \textcolor{darkgray}{85M} &\textcolor{darkgray}{87.48} &\textcolor{darkgray}{84.31} &\textcolor{darkgray}{97.31} &\textcolor{darkgray}{93.29} &\textcolor{darkgray}{90.07} \\
Linear &9/40K &75.85 &77.93 &41.78 &27.52 &55.77 \\ 
\hline \addlinespace[2pt]
BitFit &102K &86.05 &82.17 &85.51 &63.85 &79.40  \\
SPT-300 &230K &84.30 &79.73 &75.28 &40.85 &70.04  \\
DPT-25 &230K &86.52 &\cellcolor{pinksecondbest}83.67 &89.18 &68.60 &81.99  \\
Pref-T 24 &221K &82.93 &81.39 &83.46 &55.75 &75.88  \\ 
LoRA &221K &86.45 &83.83 &93.61 &76.00 &84.97\\
\hline
 \rowcolor{pastelviolet}
\multicolumn{7}{l}{\textbf{Bottleneck Adapter}}\\
\hline \addlinespace[2pt]
Pfeiffer &249K &\cellcolor{teagreen} \textbf{88.38} &83.44 &91.33 &73.19 &84.09\\\addlinespace[.05em]

Houlsby &498K &\cellcolor{pinksecondbest}88.00 &82.80 &91.75 &78.71 &85.32 \\ 

\hline
 \rowcolor{predcolor}
\multicolumn{7}{l}{\textbf{Conformer Adapter}}\\
\hline \addlinespace[2pt]
Pfeiffer &271K &88.30 &\cellcolor{teagreen} \textbf{84.57} &\cellcolor{teagreen} 96.28 &\cellcolor{pinksecondbest}95.48 &\cellcolor{teagreen} \textbf{91.16}\\ \addlinespace[.05em]

Houlsby &542K &85.97 &83.59 &\cellcolor{pinksecondbest}96.16 &\cellcolor{teagreen} \textbf{96.34} &\cellcolor{pinksecondbest} 90.51\\ 
\bottomrule

\end{tabular}

\end{table}
\endgroup

\textbf{Training Details}. For all experiments we use the AST model pre-trained on ImageNet-21K and AudioSet provided by the Huggingface Transformers library. The model has around $85.5$ million parameters, $12$ layers, and the hidden size is $768$. For the ablation studies, we also use Wav2Vec 2.0, a
well-established pre-trained model for speech tasks \cite{wav2vec}. It has around $94$M parameters and the same number of layers and hidden size as AST. For all datasets, we use AdamW optimizer with cosine annealing scheduler and weight decay set to $0.1$. The initial learning rate is $0.005$ for adapters and LoRA, while for the three prompt-tuning methods is $0.01$. Except for US8K that does not provide a validation set by default, for the others we set the hyper-parameters using the validation set. For the ESC and US8K datasets, we run 5-fold and 10-fold cross-validation as suggested in the original papers. Each experiment is carried out using a single A40/V100 GPU. The code and the complete list of hyper-parameters will be released upon acceptance.

\subsection{Main Results and Discussion}
\label{sec:main_results}

Table \ref{fig:main_diagram} presents the performance comparisons among the various PETL methods. The following observations can be drawn: \ding{182} our conformer adapter attains the best performance on average, bringing remarkable improvements over the bottleneck adapter, with the best configurations leading to up to \{4.9\%, 22.4\%\} extra performance improvement on \{GSC, FSC\}, the two datasets that exhibit the biggest mismatch between the downstream tasks and the data used for pre-training the AST model. Furthermore, our adapter approaches the FFT baseline for GSC, whereas for FSC it is capable of exceeding it by more than 3 points. \textbf{Yet, our conformer Pfeiffer/Houlsby adapter only uses 0.29/0.59\% parameters compared to the FFT baseline}. \ding{183} If we focus on the audio classification tasks, we note good improvements with respect to US8K (it also manages to outstrip FFT by 0.26 points), while for ESC-50 our adapter performs on par with the bottleneck adapter. We point out that the bottleneck adapter outperforms the FFT baseline and it can be already considered a strong approach, so using a more complex design like ours does not improve the performance accuracy. \ding{184} For the bottleneck and conformer adapters, the Houlsby configuration leads to better results for speech classification tasks, where having more parameters is beneficial (Houlsby configuration uses twice as many parameters as Pfeiffer), while for audio tasks Pfeiffer achieves better performance accuracy. \ding{185} Among the other PETL methods, we point out that LoRA achieves good results on average, beating the bottleneck Pfeiffer adapter on $3$ out of $4$ benchmarks.

\begingroup
\setlength{\tabcolsep}{2pt}
\newcolumntype{K}{!{\color{white}\ }c}

\begin{table}[t]
\centering
\caption{Few-shot analysis for the ESC-50 and GSC datasets.}
\label{tab:fewshot}
\begin{tabular}{lKKKKKKKK}
\toprule
 &\multicolumn{4}{c}{\cellcolor{pastelyellow} \textbf{ESC}} & \multicolumn{4}{c}{\cellcolor{aquamarine} \textbf{GSC}} \\
& \multicolumn{8}{c}{\textbf{Examples per class}} \\
 \cmidrule(r){2-9}
\textbf{Method} & \cellcolor{almondlow}1 & \cellcolor{almondmiddle}2 &\cellcolor{almondhigh} 4 &\cellcolor{almondultra} 8 & \cellcolor{almondlow}2 & \cellcolor{almondmiddle}8 &\cellcolor{almondhigh} 32 &\cellcolor{almondultra} 64 \\
\midrule

DPT-25 & \cellcolor{pinksecondbest}32.7 & \cellcolor{pinksecondbest}44.3 & 57.0 & \cellcolor{pinksecondbest}71.9 &\cellcolor{teagreen}\textbf{9.4} &\cellcolor{teagreen}\textbf{18.7} &43.1 &57.1 \\
LoRA &31.8 & 42.2 & \cellcolor{pinksecondbest}58.8 & 70.7 &6.8 &15.2 &41.8 &59.8\\
Bottleneck &\cellcolor{teagreen}\textbf{33.0} &\cellcolor{teagreen}\textbf{45.5} &\cellcolor{teagreen}\textbf{60.2} & \cellcolor{teagreen}\textbf{72.8} &\cellcolor{pinksecondbest}7.2 &\cellcolor{pinksecondbest}16.0 &\cellcolor{pinksecondbest}47.9 &\cellcolor{pinksecondbest}66.6 \\
Conformer & 30.7 & 41.0 & 56.2 & 71.1 & 5.9 & 15.5 & \cellcolor{teagreen}\textbf{58.7} & \cellcolor{teagreen}\textbf{77.5} \\

\addlinespace[.05em]

\bottomrule
\end{tabular}
\end{table}

\endgroup

\subsection{Ablation Studies}
\label{sec:ablations}
In this section, we study the efficacy of our proposed adapter under different settings such as few-shot learning and different pre-trained models (e.g., Wav2Vec 2.0). For the bottleneck/conformer adapters, we use the Pfeiffer configuration.

\textbf{Few-shot Analysis}. We evaluate our proposed adapters for few-shot parameter-efficient transfer learning. This scenario is challenging  because, in addition to the constraint on the number of trainable parameters, only a few samples are labeled per class. We report the accuracy results for ESC ($4$ samples) and GSC ($32$ samples) in Table \ref{tab:fewshot}. We see that, whereas for ESC the bottleneck adapter attains the best results, for GSC the gap between this and the conformer adapter is more than $10$ points. This again confirms that our proposed adapter is the best choice for speech tasks.

\begin{figure*}
\centering
\begin{subfigure}{0.35\textwidth}
    \includegraphics[width=\textwidth]{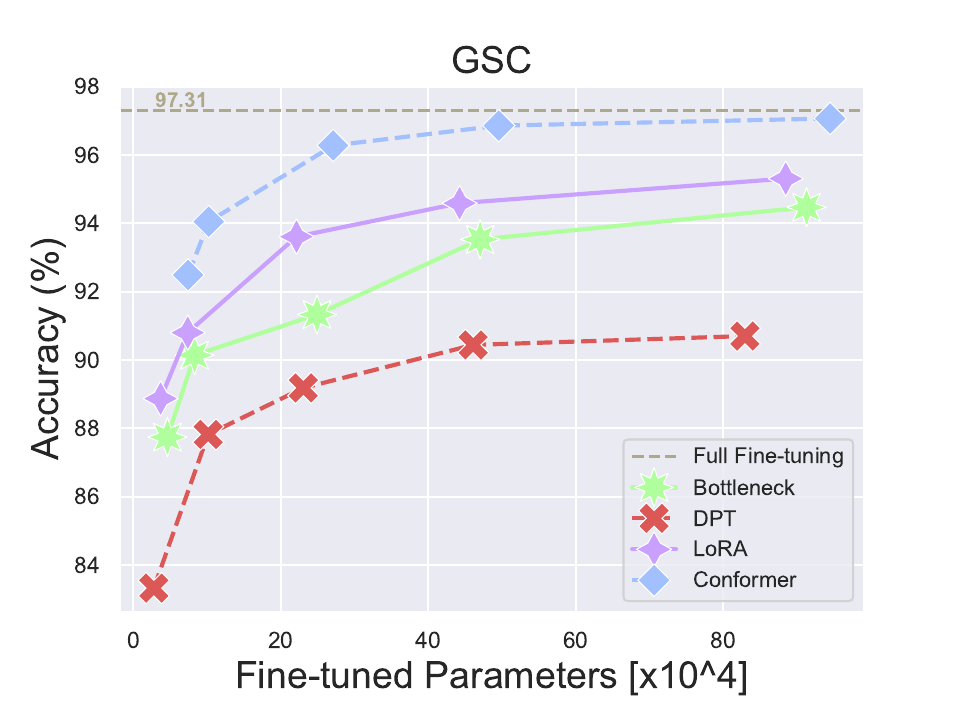}
\end{subfigure}
\begin{subfigure}{0.35\textwidth}
    \includegraphics[width=\textwidth]{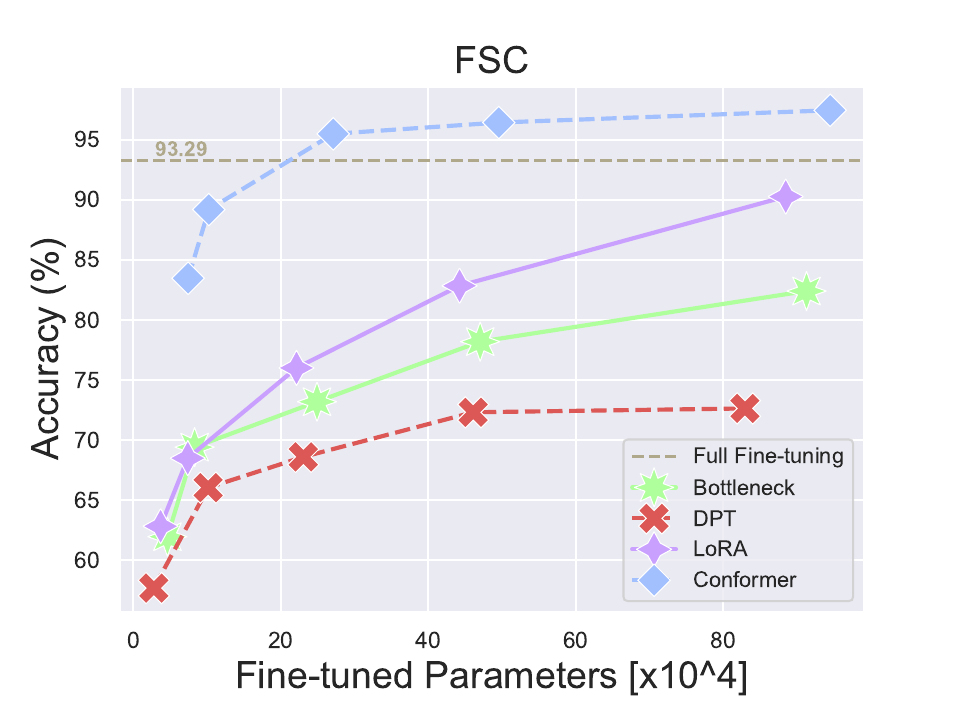}
    
\end{subfigure}
\caption{Scaling trend as more trainable parameters for each PETL method are used for GSC \textbf{(Left)} and FSC \textbf{(Right)} datasets.}
\label{fig:vary_params}
\end{figure*}

\textbf{Scaling Abilities}.
We now want to verify whether our proposed adapter also performs better when fewer parameters (e.g., $50$K) or more parameters (up to $1$M) are allocated. We restrict our analysis to the Pfeiffer configuration and GSC/FSC datasets. In Fig.~\ref{fig:vary_params} we observe that the conformer adapter, regardless of the number of parameters, outstrips the other PETL approaches. In turn, LoRA turns out to be the second best method, and it exhibits strong scaling properties when more parameters are used, bringing better results than bottleneck adapter and DPT. We point out that for FSC, the best result obtained with LoRA requires roughly $900$K parameters, whereas the conformer adapter only requires around $100$K to achieve the same accuracy.

\begin{figure}
\centering
\begin{subfigure}{0.23\textwidth}
    \includegraphics[width=\textwidth]{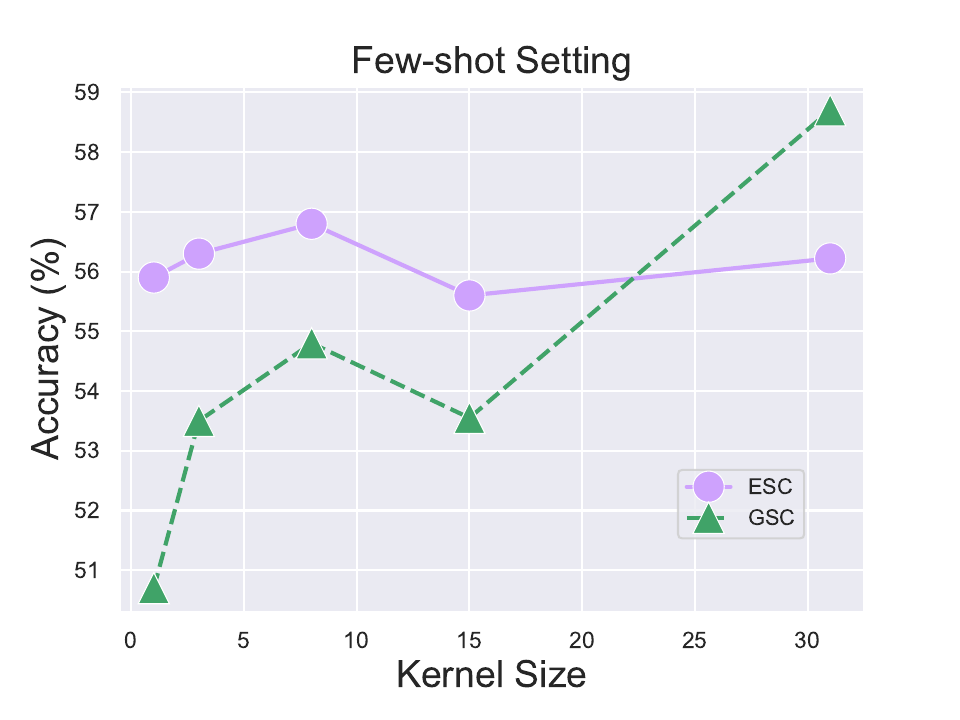}
    \label{fig:first}
\end{subfigure}
\begin{subfigure}{0.23\textwidth}
    \includegraphics[width=\textwidth]{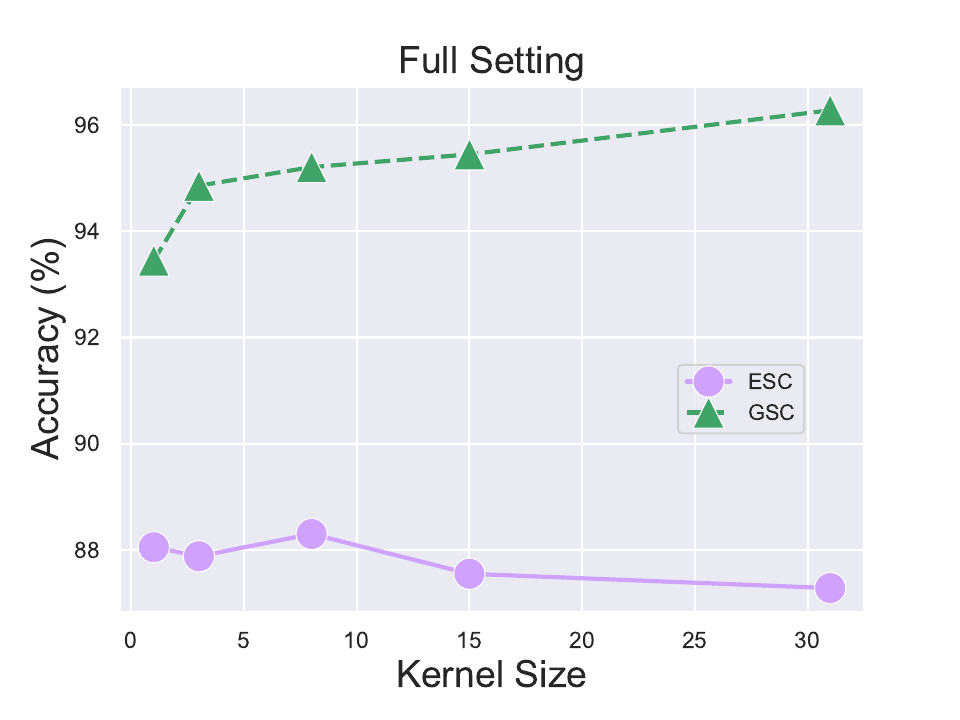}
    \label{fig:second}
\end{subfigure}
\caption{Impact of the kernel size for the ESC and GSC datasets on the few-shot \textbf{(Left)} and full \textbf{(Right)} settings.}
\label{fig:kernel_exp}
\end{figure}

\textbf{On the Kernel Size of the Conformer Adapter}. We study the impact of the kernel size $k$ of the conformer adapter on the ESC and GSC datasets both for the few-shot ($4/32$ samples) and full (i.e., no few-shot) settings. We let $k$ vary from $1$ to $31$, and we point out that setting $k = 31$ only adds roughly $2.8$K parameters with respect to $k=1$. In Fig.~\ref{fig:kernel_exp} we observe that for ESC, the results are not much influenced by $k$, and setting $k = 8$ provides the best results. On the contrary, for a more challenging dataset like GSC we see that increasing $k$ leads to better results for both the few-shot and full settings, with the former being more sensitive to it. We speculate that this happens because a larger kernel allows learning more global features, and this is more beneficial when the number of features is small (for the conformer adapter $r = 8$). On the contrary, since ESC is an easier dataset, a smaller kernel is sufficient to achieve optimal results.

\textbf{Additional Results for Wav2Vec 2.0}. 
Finally, we want to verify whether the proposed conformer adapter can be efficiently harnessed for another pre-trained model. In this direction, we consider Wav2Vec 2.0 \cite{wav2vec}. We test the bottleneck and conformer adapters on FSC and GSC, as well as IEMOCAP \cite{iemocap}, a benchmark for emotion recognition. For IEMOCAP, we increase the number of parameters to roughly $900$K as it is a more challenging dataset. As we can see from Table \ref{tab:wav2vec}, the performance gap between the adapter approaches and the FFT is large for IEMOCAP. Nonetheless, we can observe that the conformer adapter reaches accuracy results of $55.81$, with an improvement of more than $6$ points over the bottleneck adapter. For the other two datasets, the conformer adapter turns out again to surpass the bottleneck adapter.

\begingroup
\setlength{\tabcolsep}{3.5pt}
\newcolumntype{K}{!{\color{white}\ }c}

\begin{table}[t]
\centering
\caption{Results of bottlenck and conformer adapters for Wav2Vec 2.0 on GSC, FSC and IEMOCAP (IEM) datasets.}
\label{tab:wav2vec}
\begin{tabular}{lKKKKKK}
\toprule
\textbf{Method} & \textbf{Par} &\cellcolor{aqua} \textbf{GSC} &\cellcolor{melon} \textbf{FSC} & \textbf{Par} & \cellcolor{slateblue} \textbf{IEM} & \cellcolor{champagne}\textbf{Avg}\\
\midrule
\textcolor{darkgray}{FFT} & \textcolor{darkgray}{90M} &\textcolor{darkgray}{98.16} &\textcolor{darkgray}{99.58} &\textcolor{darkgray}{90M} &\textcolor{darkgray}{70.05} &\textcolor{darkgray}{89.26} \\
Linear &24K &84.93 &30.95 &4K &36.82 &50.90 \\ 
\hline \addlinespace[2pt]

\cellcolor{pastelviolet} \textbf{Bottle} &250K &94.96 &96.41 &895K &48.32 &79.90\\ \addlinespace[.05em]

\cellcolor{predcolor} \textbf{Conf} &272K &\cellcolor{teagreen} \textbf{95.24}&\cellcolor{teagreen} \textbf{98.33} &927K &\cellcolor{teagreen} \textbf{55.81} & \cellcolor{teagreen}\textbf{83.13}\\\addlinespace[.05em]

\bottomrule
\end{tabular}
\end{table}

\endgroup

\section{Conclusion}

In this work, we study the problem of parameter-efficient
transfer learning for the AST model. To do so, we establish a framework that allows us to examine the performance achieved by the most common PETL methods across five audio/speech benchmarks. We also propose a new adapter module that relies on the conformer convolution module, making effective use of the depthwise convolution. We show that our proposed adapter turns out to be competitive with the full fine-tuning approach and outperforms the established bottleneck adapter, as well as LoRA and prompt-tuning methods. The conformer adapter also provides strong results under few-shot settings, when we vary the number of parameters and if applied to another pre-trained model like Wav2Vec 2.0. Finally, we study the role of kernel size, underscoring its pivotal role in achieving peak performance.

\bibliographystyle{IEEEbib}
\bibliography{refs}

\end{document}